# Effect of local environment on moment formation in iron silicides.

N.G. Zamkova, V.S. Zhandun, S.G. Ovchinnikov, and I.S. Sandalov

*Kirensky Institute of Physics, Siberian Branch of Russian Academy Sciences, 660036 Krasnoyarsk, Russia*

The effect of local environment on the formation of magnetic moments on *Fe* atoms in iron silicides are studied by combination of *ab initio* and model calculations. The suggested model includes all *Fe d-* and *Si p*-orbitals, intra-atomic Coulomb interactions, inter-atomic *Fe-Fe* exchange and hopping of electrons to nearest and next nearest neighboring atoms and takes into account all symmetries within the Slater-Koster scheme. The parameters of the model are found from the requirement that *self-consistent* moments on atoms and density of states found from *ab initio* and model calculations within the Hartree-Fock approximation are close to each other as much as possible. Contrary to the commonly accepted statement that an increase of the *Si* concentration within nearest environment of *Fe* atoms in the ordered *Fe$_3$Si* and *Fe$_x$Si$_{1-x}$* alloys leads to a decrease of *Fe* magnetic moment we find that a crucial role in the formation of magnetic moments is played by second coordination sphere of *Fe* atoms. Particularly, the *Fe* atoms have higher magnetic moments in amorphous films compared to the epitaxial ones due to decrease in the number of iron-atoms' in the next nearest environment. Both our model and *ab initio* calculations confirm existence of known spin crossover with pressure and predict second crossover at higher pressure.

## 1. Introduction

The transition metal silicides, particularly iron silicides, offer a large variety of potential spintronic, microelectronic and optoelectronic applications for silicon-based devices because, depending on their phase, crystal structure and composition, they can be semiconducting or metallic with different ferro- (FM), ferri- (FiM) or paramagnetic (PM) states. The binary Heusler alloy, *Fe$_3$Si*, is a potentially good candidate for a spin injector. This material has a high Curie temperature (≈840 K) and has been theoretically shown to possess high spin-polarization [1-5]. At the *Fe*-rich side of the binary phase diagram, metallic as well as ferromagnetic *Fe$_5$Si$_3$* and *Fe$_3$Si* have already been established as key materials for spintronics. The *Si*-rich side of the phase diagram [6] contains several variants of a disilicide stoichiometric compound, such as the high-temperature tetragonal metallic *α-FeSi$_2$* phase, with applications as an electrode or an interconnect material, and the orthorhombic semiconducting *β-FeSi$_2$* phase, which due to its direct band gap is an interesting candidate for thermoelectric and optoelectronic devices. One of the motivations for studying the *Fe – Si* system is the possibility to tune its magnetic properties. The experiments [7,8] on bulk *Fe$_x$Si$_{1-x}$* alloys show that the magnetic properties are strongly depend on *Si* concentration and chemical order. The local magnetic moments at *Fe* sites may become higher than in pure iron, depending on the distribution of *Fe* and *Si* neighbors, and disappear at *Si* concentration of nearly 50%. The presence of the *Si* neighbors decreases the average magnetic moment at the *Fe* sites, resulting in the appearance of high- and low-spin *Fe* species. This has been studied for ordered and disordered *Fe$_x$Si$_{1-x}$* by neutron diffraction [9,10], Mössbauer effect measurements [11,12] and pulsed NMR studies [13]. These iron silicides are also technologically advantageous since they can be grown epitaxially on many different semiconductor and insulator substrates [14-19]. What makes the system *Fe$_x$Si$_{1-x}$* unique is that it allows for varying the degrees of both chemical and structural order over a wide composition range with the thin film growth techniques; the high-quality epitaxial films on *Si* may exhibit ferromagnetism. The latter promises perspective for the integration of the *FeSi*-based magnetic devices into silicon technology. Furthermore, the iron silicides, which do not exist in bulk, can be stabilized as films. Recently a successful fabrication of thin films solid solution *Fe$_x$Si$_{1-x}$* within the composition range 0.5<*x*<0.75 with the *CsCl* structure (B2) was reported [14,16]. Also, while the magnetic order is not observed in bulk stoichiometric disilicide *FeSi$_2$*, ferromagnetism was found [20] in the metastable phase α-FeSi2, which was stabilized in epitaxial-film form on the silicon substrate.

Most of theoretical works were devoted to the solid solutions *Fe-Si* from *Fe*-rich side of the phase diagram with *bcc*-like structures (DO$_3$, B2, A2). The phenomenological models [11,21,22], have been suggested for the explanation of the measured hyperfine fields at atoms *Fe* with different numbers of *Fe* as first neighbors. They obtained consistent with experiment linear decrease of the *Fe* moment with concentration of *Si* ions in first coordination sphere. The electron structure and magnetic properties of the ordered bulk *Fe$_3$Si* and based on its solid solutions *Fe$_x$Si$_{1-x}$*, *Fe$_{3-x}$T$_x$Si*, where *T* is a transition metal, *Fe$_{3-x}$V$_x$X* with *X=Si,Ga,Al*, studied by *ab initio* calculations [23-27]. The calculation of moments and electronic structure of binary *Fe$_x$Si$_{1-x}$* and ternary *Fe$_{3-x}$V$_x$Si* random alloys with DO$_3$-like structure within coherent potential approximation (CPA) [23] has confirmed the conclusions of the phenomenological model about practically linear variation of the *Fe* magnetic moment with the number of nearest *Fe* neighbors. The same was stated in the later theoretical works [24-27]. However, as emphasized in the works [8, 28] on Mössbauer spectra, the contribution of the second neighbors of *Fe* ions to the formation of its moment is far from being negligible. The non-cluster CPA, by construction being an effective-medium method, as well as local environment models [11,21,22], provide information on influence of average number of metallic or metalloid atoms on the magnetic moment formation (MMF). For this reason they are not able to provide information about the role played by the different local environment at the same concentration of the alloy components. We found two theoretical attempts to attract attention to this problem [29,30]. In first work [29] the *d*-electrons were mimicked by Hubbard's *s*-band, hybridized with the *Si* single-band of non-interacting electrons; the dependence on local environment was described via the position of *d*-level. The authors [29] came to the conclusion that the local magnetic moment is determined by the number of metalloid atoms and weakly depends on the concentration. Due to oversimplification of the model it remains unclear, however, if these conclusions are related to the compounds of interest or not. More realistic model [30], which is close to the model, which we will use here, includes all five *3d*-electron orbitals of *Fe* and three *3p*-orbitals *Si*, and uses the Slater-Koster approach [31] for hopping integrals. The work [30] was devoted to the solution of the experimentally known puzzle: why the impurities from left side of Fe (say, Mn) prefer to occupy the cubic-symmetry sites, whereas those from right side (like Co) the tetrahedral-symmetry sites. The magnetism in this work was treated via the only Stoner's exchange-splitting parameter, which was used to fit the *average* magnetic moment to the experimental one. Unfortunately, the latter simplification does not allow to describe the effects of local environment in these compounds.

The target of this work is to investigate the influence of local environment on the formation of the iron-atoms' magnetic moments in iron silicides *Fe$_3$Si* and solid solutions *Fe$_x$Si$_{1-x}$*. Particularly, we will address the question, raised in the experimental works [8,28] about the role, played by second neighbors of *Fe* ions in the physics of MMF on it. The dependence of the *Fe* moments on pressure and the possibility of high-spin - low-spin crossover also will be investigated.

The paper is organized as follow. In Sec.2 we formulate the multiorbital model and provide the details of *ab initio* calculations. In Sec.3 the results of the model and *ab initio* calculations of *Fe$_3$Si* and its alloys are compared and the dependence of magnetic moments on the hopping matrix elements is presented. The spin-crossovers under pressure in *Fe$_3$Si* are described in Sec.4. The formation of magnetic moments in the alloys *Fe$_x$Si$_{1-x}$* is considered in Sec.5. Sec.6 contains the summary of the obtained results and conclusions.

2. **The approach**

It is difficult to separate the contribution of the second neighbors of *Fe* ions to the MMF on it. For this reason we combine the *ab initio* calculations with the model one. We use the following scheme. First we perform the calculation of electronic and magnetic properties of the compound of interest within the framework of density functional theory in the generalized gradient approximation (DFT-GGA) for positionally different substitutions of silicon atoms by the iron-atoms'. Then we perform mapping the DFT-GGA results to the model. The guiding arguments for the formulation of the model are: the model should 1) contain as little as possible parameters; 2) contain the specific information about the compound in question (*i.e..*, contain the proper number of orbitals and electrons and reflect the crystal structure) and 3) contain main interactions, reflecting our understanding of the underlying physics. At last, we perform the mapping following the DFT ideology: we find the parameters of the model from fitting *its self-consistent charge density* to the one, obtained by *ab initio* calculations. The latter step distinguishes our approach from the other ones.

**2.1 The *ab initio* part**

All *ab initio* calculations presented in this paper are performed using the Vienna *ab initio* simulation package (VASP**)** [32] with projector augmented wave (PAW) pseudopotentials [33]. The valence electron configurations $3d^64s^2$ are taken for *Fe* atoms and $3s^23p^2$ for *Si* atoms. The calculations are based on the density functional theory where the exchange-correlation functional is chosen within the Perdew-Burke-Ernzerhoff (PBE) parameterization [34] and the generalized gradient approximation (GGA) has been used. Throughout all calculations, the plane-wave cutoff energy 500 eV is used. The Brillouin-zone integration is performed on the grid Monkhorst-Pack [35] special points 10×10×10.

**2.2 The model part**

One can conclude from the experiments [9-13] that the *d*-electrons in iron silicides are delocalized. At the same time there is a consensus in scientific community that the intraatomic interactions are strong enough to contribute to formation of the moment on *Fe* ions. We include into the Hamiltonian of our model this set of interactions between the *d*-electrons of *Fe* (5 *d*-orbitals per spin) following Kanamori [36]. The structure contains neighboring *Fe* ions, therefore, the interatomic direct *d-d*-exchange and *d-d*-hopping cannot be ignored. The *Si p*-electrons (*3p*-orbitals per spin) are modeled by atomic levels and interatomic hoppings, no *p-p*-Coulomb terms are included. Both subsystems are connected by *d-p*-hoppings. Thus, the Hamiltonian of the model is:

$$H = H^{Fe} + H_{J'}^{Fe-Fe} + H_0^{Si} + H_{hop}, \quad H^{Fe} = H_0^{Fe} + H_K^{Fe} \tag{1}$$

where

$$H_0^{Fe} = \sum \varepsilon_0^{Fe}\, \hat{n}_{nm\sigma}^d; \quad H_0^{Si} = \sum \varepsilon_0^{Si}\, \hat{n}_{nm\sigma}^p;$$

and the Kanamori's part of the Hamiltonian

$$H_K^{Fe} = \frac{U}{2}\sum \hat{n}_{nm\sigma}^d \hat{n}_{nm\bar{\sigma}}^d + \left(U' - \frac{1}{2}J\right)\sum \hat{n}_{nm}^d \hat{n}_{nm'}^d (1 - \delta_{mm'}) - \frac{1}{2}J \sum \hat{s}_{nm}^d \hat{s}_{nm'}^d;$$

$$H_{J'}^{Fe-Fe} = -\frac{1}{2}J' \sum \hat{s}_{nm}^d \hat{s}_{n'm'}^d;$$

$$\mathcal{H}_{hop} = \sum T_{nn'}^{mm'} p_{nm\sigma}^\dagger p_{n'm'\sigma} + \sum t_{nn'}^{mm'} d_{nm\sigma}^\dagger d_{n'm'\sigma} + \sum [(t')_{nn'}^{mm'} d_{nm\sigma}^\dagger p_{n'm'\sigma} + H.c.];$$

$$\hat{n}_{nm\sigma}^d \equiv d_{nm\sigma}^\dagger d_{nm\sigma}; \quad \hat{n}_{nm}^d = \hat{n}_{nm\uparrow}^d + \hat{n}_{nm\downarrow}^d; \quad \hat{s}_{nm}^d \equiv \sigma_{\alpha\gamma} d_{nm\alpha}^\dagger d_{nm\gamma}; \quad \hat{n}_{nm\sigma}^p \equiv p_{nm\sigma}^\dagger p_{nm\sigma}. \tag{2}$$

Here $p_{nm\sigma}^\dagger (p_{nm\sigma})$ and $d_{nm\sigma}^\dagger (d_{nm\sigma})$ are the creation (annihilation) operators of *p*- electrons of *Si*- and *d*-electrons of *Fe*-ions; *n* - complex index lattice (site, basis); *m* – label the orbital; $\sigma$ is spin projector index; $\sigma_{\alpha\gamma}$ are Pauli matrices; $U, U' = U - 2J$ and $J$ are the intra-atomic Kanamori parameters; $J'$ is the parameter of the intersite exchange between nearest *Fe* atoms. At last, $T_{nn'}^{mm'}, t_{nn'}^{mm'}, (t')_{nn'}^{mm'}$ are hopping integrals between atoms *Si-Si*, *Fe-Fe* and *Fe*-S atoms, correspondingly. Notice that in order to reduce the number of parameters we did not include into Hamiltonian the terms, describing the crystal electric field. Therefore, in atomic limit (all *t*=0) all iron-atoms' are completely identical. Since the hopping matrix elements fully reflect the crystal symmetry, they provide the splitting of the atomic states of *Fe* ions according to the symmetry of the local environment. They are fitting parameters of the model; for this reason we avoid to introduce additional crystal-field parameters.

Since our target is to obtain the zero-temperature phase diagram for magnetic moment formation and in present report we do not study the thermodynamics, it is sufficient to analyze the system within Hartree-Fock approximation (HFA) from weak-coupling side. After standard HFA decoupling and Fourier transformation we obtain the matrix equation for the Green's functions for each spin for spin-homogeneous states:

$$\left(E\delta_{ii_1}\delta_{mm_1} - \Omega_{im,i_1m_1}^\sigma(\boldsymbol{k})\right) G_{i_1m_1,i'm'}^\sigma(\boldsymbol{k}, E) = \delta_{ii'}\delta_{mm'}, \tag{3}$$

where *m* stands for orbital and *I* labels atom in the basis. The matrix $\Omega_{\mathfrak{I},i_1m_1}^\sigma$ consist of the blocks,

$$\Omega(k) = \begin{pmatrix} \Omega_{FeFe}(k) & \Omega_{FeSi}(k) \\ \Omega_{FeSi}^\dagger(k) & \Omega_{SiSi}(k) \end{pmatrix}, \tag{4}$$

which have the form

$$[\Omega_{FeFe}(\boldsymbol{k})]^{\sigma}_{\lambda i,\mu j} = \delta_{\lambda\mu}\delta_{ij}\varepsilon^{Fe}_{i\sigma} - t^{\lambda\mu}_{ij}(\boldsymbol{k}), \quad [\Omega_{FeSi}(\boldsymbol{k})]^{\sigma}_{\lambda i,\mu j} = [t'(\boldsymbol{k})]^{\lambda\mu}_{ij}, \tag{5}$$

$$[\Omega_{SiSi}]^{\sigma}_{\lambda i,\mu j}(k) = \delta_{\lambda\mu}\delta_{ij}\varepsilon^{Si}_{i\sigma} - T^{\lambda\mu}_{ij}(\boldsymbol{k}) \tag{6}$$

with

$$\varepsilon^{Fe}_{i\sigma} = \varepsilon^{Fe}_0 + U n^{d,\bar{\sigma}}_{i\lambda} + U' \sum_{m\neq\lambda} n^d_{im} - 2J\eta(\sigma)\sum_m \sigma^{d,z}_{im} - 2J'\eta(\sigma)\sum_{l,m}\sigma^{d,z}_{lm} \tag{7}$$

$$n^{d,\bar{\sigma}}_{i\lambda} = \sum_k \langle d^{\dagger}_{k,i\lambda\bar{\sigma}} d_{k,i\lambda\bar{\sigma}}\rangle, \quad n^{d,\sigma}_{i\lambda} = \sum_k \langle d^{\dagger}_{k,i\lambda\sigma} d_{k,i\lambda\sigma}\rangle, \quad \sigma^{d,z}_{im} = \tfrac{1}{2}\sum_\sigma \sigma n^{d,\sigma}_{im}. \tag{8}$$

And $\eta(\uparrow) = 1, \eta(\downarrow) = -1$. Then the self-consistent equations for population numbers are expressed in terms of eigenvalues $\varepsilon^{\sigma}_{\nu}(\boldsymbol{k})$ and eigenvectors $u^{\nu\sigma}_{im}(\boldsymbol{k})$ of matrix $\Omega^{\sigma}_{im,i_1m_1}(\boldsymbol{k})$:

$$n^{d,\sigma}_{im}(\boldsymbol{k}) = \langle d^{\dagger}_{k,im\sigma} d_{k,im\sigma}\rangle = \sum_\nu [u^{\nu\sigma}_{im}(\boldsymbol{k})]^* f(\varepsilon^{\sigma}_{\nu}(\boldsymbol{k}) - \mu) u^{\nu\sigma}_{im}(\boldsymbol{k}). \tag{9}$$

Particularly, for the $Fe_3Si$ the matrix $\hat{\Omega}$ has size 18×18 for each spin. The function $f(x) = [1 + exp\,(x/T)]^{-1}$ is Fermi function, chemical potential $\mu$ is found, as usual, from the full number of electrons per the cell. The dependences of hopping integrals $t^{mm'}_{nn'}$, $(t')^{mm'}_{nn'}$, $T^{mm'}_{nn'}$ on $\boldsymbol{k}$ were obtained from the Slater and Koster atomic orbital scheme [31] in the two-center approximation using basic set consisting of five $3d$ orbitals for each spin on each $Fe$ and three $3p$ orbital for each spin on each $Si$. In this *two-centre approximation* the hopping integrals depend on the displacement $\boldsymbol{R}=(l\mathbf{x}+m\mathbf{y}+n\mathbf{z})$ between the two atoms, where $\mathbf{x, y, z}$ are the unit vectors along cubic axis and $l, m, n$ are direction cosines. Then, within the two-center approximation, the hopping integrals are expressed in terms of Slater–Koster parameters $t_\sigma\equiv(dd\sigma)$, $t_\pi\equiv(dd\pi)$ and $t_\delta\equiv(dd\delta)$ for $Fe-Fe$ hopping and $t_\sigma\equiv(pd\sigma)$, $t_\pi\equiv(pd\pi)$ for $Fe-Si$ and $Si-Si$ hoppings ($\sigma, \pi, \delta$ specifies the component of the angular momentum relative to the direction $\boldsymbol{R}$). Their $\boldsymbol{k}$-dependence are given by the functions $\gamma_\sigma(\boldsymbol{k})$, $\gamma_\pi(\boldsymbol{k})$ and $\gamma_\delta(\boldsymbol{k})$, where $\gamma(\boldsymbol{k}) = \sum_R e^{ikR}$. The expressions for hopping integrals can be obtained with Table I from [31]. For example, $t^{xy,xy}_{Fe-Fe}(\boldsymbol{k}) = 2t_\pi(cos(R_x k_x)+cos(R_y k_y))+2t_\delta cos(R_z k_z)$, etc.

The population numbers $n^{d,\sigma}_{im}(\boldsymbol{k})$, $n^{p,\sigma}_{im}(\boldsymbol{k})$ have been found self-consistently with the accuracy $o(10^{-3})$. The number of point in the Brillouin zone for the $FCC$ lattice was taken 512 and 1000 for the $SC$ lattice. The Monkhorst-Pack scheme [35] was used for generation of the k-mesh. The calculations were performed from three initial states: $FM$, $AFM$ and $PM$ states. After achieving self-consistency the state with minimal total energy was chosen. The last step was done with the help of the Galitsky-Migdal formula for total energy, which we adopted for our model. Within HFA it acquires the form

$$E_{tot} = \tfrac{1}{2}\sum_{ij\lambda\mu}\sum_{\nu k}[u^{\nu\sigma}_{im}(\boldsymbol{k})]^*\left(\tau^{\lambda\mu}_{ij}(\boldsymbol{k}) + \varepsilon^{\sigma}_{\nu}(\boldsymbol{k})\right) f(\varepsilon^{\sigma}_{\nu}(\boldsymbol{k}) - \mu) u^{\nu\sigma}_{im}(\boldsymbol{k}), \tag{10}$$

with

$$\tau^{\lambda\mu}_{ij}(\boldsymbol{k}) = \delta_{ij}\delta_{\lambda\mu}\varepsilon^0_i + t^{\lambda\mu}_{ij} + (t')^{\lambda\mu}_{ij} + T^{\lambda\mu}_{ij} \tag{11}$$

(we remind that the basis indices $i,j$ label the sorts of atoms)
.

### 3. The dependence of moments on hopping.

Stoichiometric $Fe_3Si$ has the $DO_3$ crystal structure with the space symmetry group $Fm\bar{3}m$ and with four atoms in the elementary cell: one $Fe_I$, two $Fe_{II}$ и $Si$ (Fig.1a). The iron-atoms' have two nonequivalent crystallographic positions $Fe_I$ and $Fe_{II}$. $Fe_I$ and $Si$ sit in cubic positions with the point-group symmetry $O_h$, whereas $Fe_{II}$ are in tetrahedral positions with symmetry $T_d$. Inequivalent $Fe$ ions have different local environment in both first (NN) and second coordination spheres (NNN). $Fe_I$ is surrounded by eight $Fe_{II}$ in NN, six $Si$ ions in NNN, whereas NN of $Fe_{II}$ contains four $Fe_I$ atoms and four $Si$, while NNN of $Fe_{II}$ consists of six $Fe_{II}$ atoms. Such a different distribution of $Si$ neighbors leads to very different magnetic moments μ on ions of $Fe$: μ($Fe_I$)=2.5μ$_B$ and μ($Fe_{II}$)=1.35μ$_B$ [21,23].

All *ab initio* calculations of the density of electron states (DOS) and magnetic moments are performed on the equilibrium lattice parameters, which are found from the full optimization of the structure geometries within GGA. For the $Fe_3Si$ in $DO_3$-type of structure it is $a = 5.60$Å. The distance between neighbors in NN are $R_{NN} = 2.42$Å and in NNN are $R_{NNN} = 2.80$Å. The nearest neighbors of $Si$ atoms are separated by $R_{Si-Si} = a\frac{\sqrt{2}}{2}$ and are third neighbors.

In all model calculations we used the following parameters: $U=1$ (i.e. all other parameters are units of $U$), $J=0.4$, $J'=0.05$, $\varepsilon^{Si}=6$, $\varepsilon^{Fe}=0$. There is five hopping parameters for the $Fe_3Si$: $t_1$ and $t_2$ in NN, $t_1 \equiv t(R_{FeI-FeII})$ for $Fe_I$ - $Fe_{II}$ and $t_2 \equiv t(R_{FeII-Si})$ for $Fe_{II}$ – $Si$; $t_3$, $t_4$ in NNN, $t_3 \equiv t(R_{FeII-FeII})$ for $Fe_{II}$-$Fe_{II}$ and $t_4 \equiv t(R_{FeI-Si})$ for $Fe_I$ – $Si$ for NNN; and $t_5 \equiv t(R_{Si-Si})$ for Si –Si. In order to decrease the number of parameters, the weak δ-bonds are neglected ($t_{1\delta}=t_{3\delta}=0$).

The values of the parameters are found from the requirement that after achieving self-consistency in both the model (HFA) and in *ab initio* calculations (GGA), the *d*-DOS and magnetic moments on *Fe* atoms have to be as close to each other as possible. The comparison of HFA and GGA *Fe d*-population numbers are shown in the Table 1 at the parameters, shown in last column at the Table 1. As seen, these parameters provide small enough error in occupations of the orbitals to believe that the model reflects the properties of real compounds.

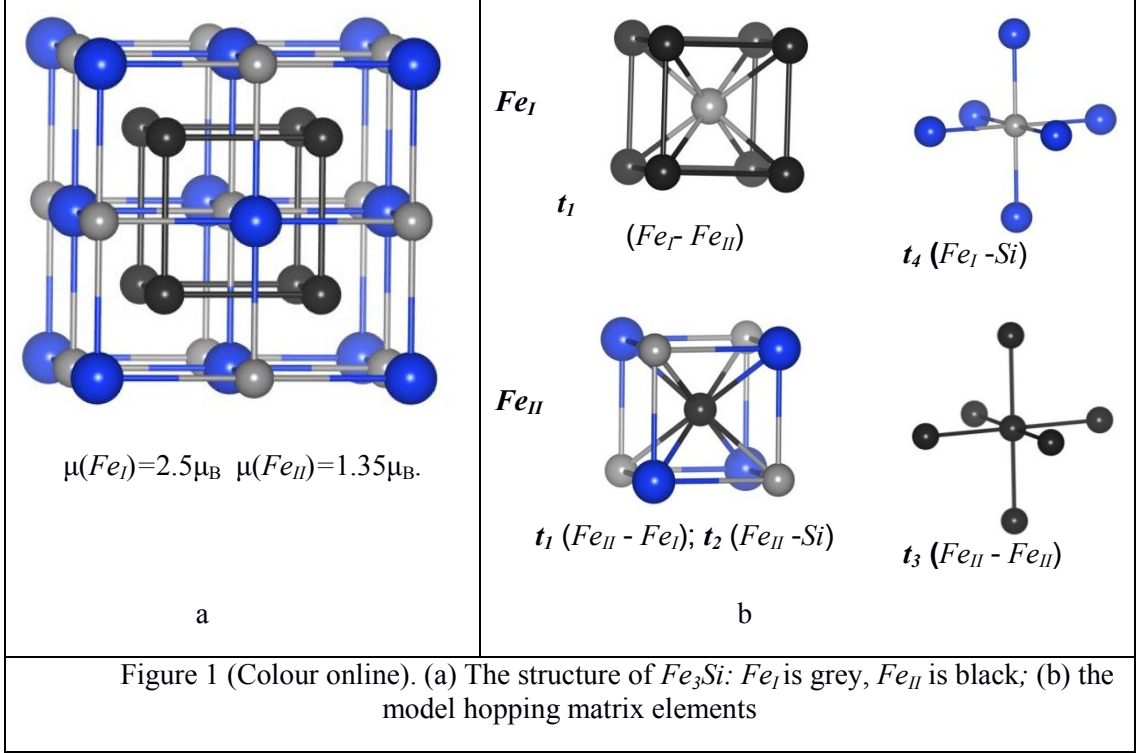

Figure 1 (Colour online). (a) The structure of $Fe_3Si$: $Fe_I$ is grey, $Fe_{II}$ is black; (b) the model hopping matrix elements

The corresponding partial DOS (pDOS) are compared in Fig. 2. As seen, qualitatively the features of the *ab initio* pDOS, namely, the peculiarities *d*- pDOS for inequivalent *Fe* atoms, like well developed peaks for $Fe_I$ and smeared pDOS for $Fe_{II}$, are reproduced by the model. However, the model pDOS occupies narrower interval of energy and is concentrated in the region closer to Fermi energy. As mentioned above, we assume that this is due to absence of *s*-, *p*-electrons of *Fe* and *s*-electrons of *Si* in the model.

**Table 1.** The best fit of the model parameters in HFA to GGA-DFT orbital populations ($n_\uparrow^d, n_\downarrow^d$), magnetic moments (μ) and the number of electrons ($N_{el}$)

|  | Orbital | VASP | | | | Model | | | | Hopping parameters |
|---|---|---|---|---|---|---|---|---|---|---|
|  |  | $n_\uparrow^d$ | $n_\downarrow^d$ | μ, μ$_B$ | $N_{el}$ | $n_\uparrow^d$ | $n_\downarrow^d$ | μ, μ$_B$ | $N_{el}$ | $t_{1\sigma}=0.55$  $t_{1\pi}=t_{1\sigma}/3=0.187$ |
| $Fe_I$ | $t_{2g}$ | 0.84 | 0.51 | 2.52 | 6.2 | 0.82 | 0.43 | 2.55 | 6.1 | $t_{2\sigma}=1.0$  $t_{2\pi}=t_{2\sigma}/2=0.5$ |
|  | $e_g$ | 0.92 | 0.18 |  |  | 0.94 | 0.25 |  |  | $t_{3\sigma}=0.4$  $t_{3\pi}=t_{3\sigma}/2=0.2$ |
| $Fe_{II}$ | $t_{2g}$ | 0.79 | 0.57 | 1.32 | 6.4 | 0.79 | 0.53 | 1.35 | 6.5 | $t_{4\sigma}=0.7$  $t_{4\pi}=t_{4\sigma}/2=0.35$ |
|  | $e_g$ | 0.77 | 0.38 |  |  | 0.78 | 0.49 |  |  | $t_{5\sigma}=0.8$  $t_{5\pi}=t_{5\sigma}/2=0.4$ |

The relation between $t_\sigma$ and $t_\pi$ shown in Table 1 was kept in all model calculations, for this reason further everywhere we will use $t_\sigma \equiv t$.

Let us first consider first the picture of the magnetic moment formation on *Fe* ions suggested in earlier model [11,21,22] and CPA [23] calculations for $Fe_3Si$ when it is fully determined by the NN environment. Corresponding *Fe*-moment dependences on the NN hopping constants $t_1$ and $t_2$ at NNN $t_3=t_4=0$ is shown in Fig. 3. As seen, the ferromagnetic solution exists at any considered values $t_1$, $t_2$ and the magnetic moments at both type of $Fe$ ions weakly depend on the $t_2$ NN $Si$- $Fe_{II}$ hopping being determined mainly by the $t_1$, NN $Fe_I$ - $Fe_{II}$. The dependence of the moment $\mu_{FeI}(t_1)$ sharper than $\mu_{FeII}(t_1)$. Particularly, the moment $\mu_{FeI}(t_1)$ drops quite fast to zero near the line $|t_1|\approx 0.7$, whereas $\mu_{FeII}(t_1)$ dependence remains smooth. What is important, however, the values of moments close to either experimental or GGA values, do not exist (there is $\mu_{FeI} \approx 2.5\mu_B$ in the narrow region near the line $|t_1|\approx 0.7$, but then $2\mu_B < \mu_{FeII}(t_1) < 3\mu_B$ and never achieves the value about $1.5\mu_B$).

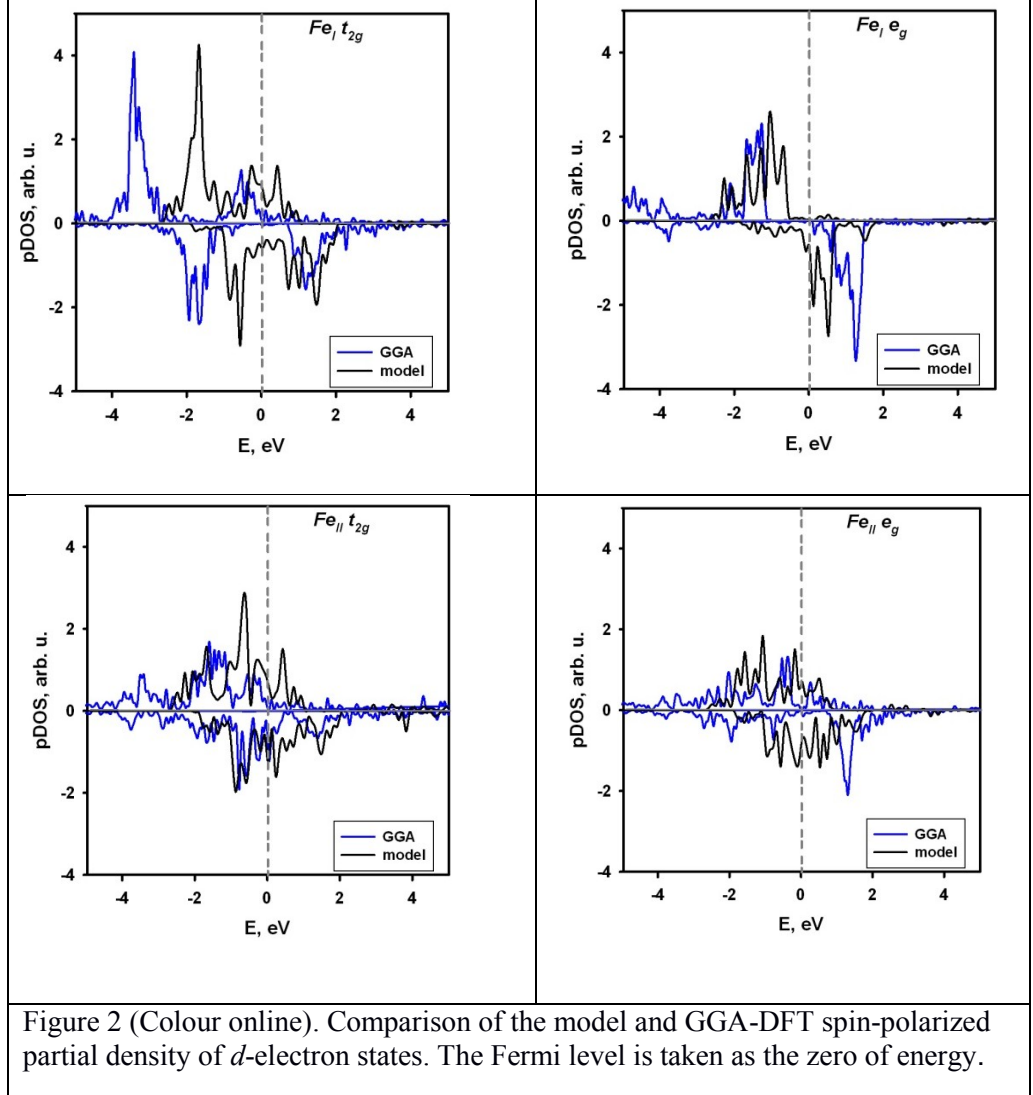

Figure 2 (Colour online). Comparison of the model and GGA-DFT spin-polarized partial density of *d*-electron states. The Fermi level is taken as the zero of energy.

Now let us switch on the NNN hoppings and build the moment maps at the fixed values $t_3(Fe_{II}$-$Fe_{II})$=0.4 and $t_4(Fe_I$-$Si)$=0.7. This results in two new regions with $2\mu_B < \mu_{FeI}(t_1) < 3\mu_B$ and $1\mu_B < \mu_{FeII}(t_1) < 2\mu_B$ (see Fig.3b). This explicitly shows that the role played by the NNN interactions is critically essential for the formation of realistic $Fe$ moments in $Fe_3Si$. The sensitivity of moments to $t_3(Fe_{II}$-$Fe_{II})$ is much higher than to $t_4(Fe_I$-$Si)$. Its increase makes the region of ferromagnetic state (FM) more narrow (Fig.3c) and at $|t_1|>0.5$ the region with ferrimagnetic state (FiM) of the kind $Fe_{II}\uparrow$-$Fe_I\downarrow$ with $|\mu_{FeI}|>|\mu_{FeII}|$ arises. Besides, namely the NNN hopping $t_3(Fe_{II}$-$Fe_{II})$ leads to the delocalization of *d*-electrons of $Fe_{II}$, the decrease of $\mu_{FeII}$ and an increase of the pDOS of *d*-electrons of $Fe_{II}$. The switching on the NNN hopping $t_4(Fe_I$-$Si)$ flips the moment $\mu_{FeII}$ and the region with ferromagnetic state disappears. However, it worth noting that the FM state in $Fe_3Si$ is stabilized by the intersite exchange interaction $J'$. Indeed, at $J'$=0 the FiM state is stable in all considered regions of the ($t_1,t_2,t_3,t_4$)-space.

In order to get a hint why the moment is so sensitive to the hopping $t_3(Fe_{II} - Fe_{II})$, we can simplify the problem till the analytically solvable level. Namely, we replace the each many-orbital block by single-orbital one in the secular equation matrix:

$$\begin{array}{c} \phantom{Fe_I}\begin{array}{ccc} Fe_I & Fe_{II} & Fe_{II} \end{array} \\ \begin{array}{c} Fe_I \\ Fe_{II} \\ Fe_{II} \end{array} \left| \begin{array}{ccc} \varepsilon_{Fe_I} & t_1(k) & t_1(k) \\ t_1(k) & \varepsilon_{Fe_{II}} & t_3(k) \\ t_1(k) & t_3(k) & \varepsilon_{Fe_{II}} \end{array} \right| \end{array} \quad (12)$$

The eigenvalues of this matrix are

$$E_1(\boldsymbol{k}) = \varepsilon_{Fe_I} - t_3(k);$$

$$E_{2,3}(\boldsymbol{k}) = \frac{1}{2}\left[\varepsilon_{Fe_I} + \varepsilon_{Fe_{II}} \pm \sqrt{\left(\varepsilon_{Fe_I} - \varepsilon_{Fe_{II}} - t_3(k)\right)^2 + 8t_1^2(k)}\right] \quad (13)$$

As seen, if the hopping $t_3(k)$ vanishes, the energy $E_1(\boldsymbol{k})=\varepsilon_{Fe_I}$, i.e., becomes atomic level. Therefore, in this limit this state acquires the atomic magnetic moment. Of course, in real multi-orbital case the situation is more complex and fully atomic solution does not appear, but, as follows from the solution of the full problem the tendency remains the same.

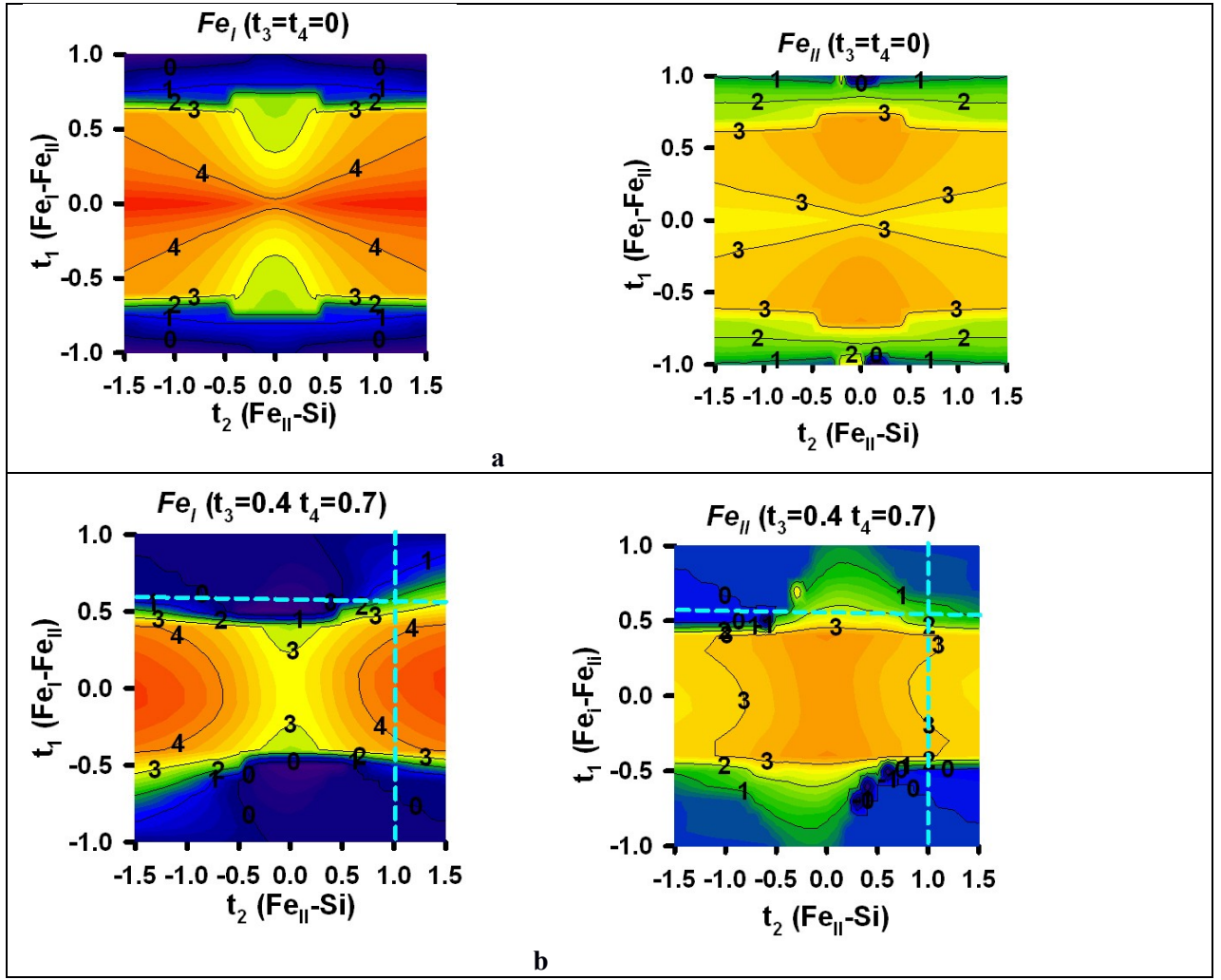

a

b

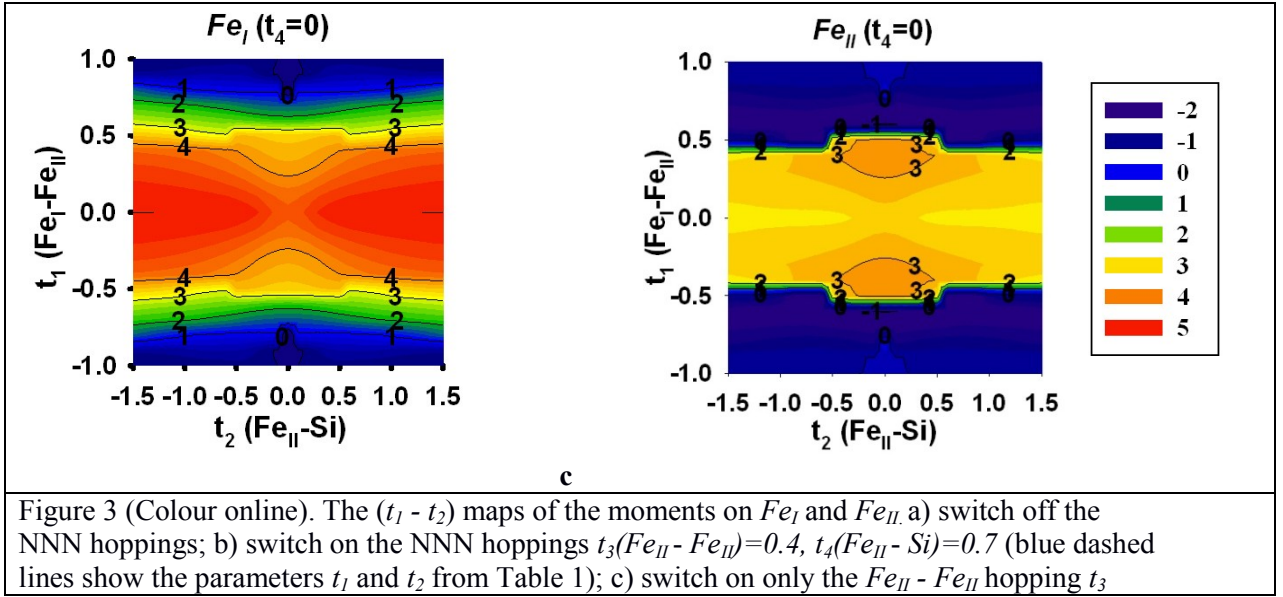

Figure 3 (Colour online). The ($t_1$ - $t_2$) maps of the moments on $Fe_I$ and $Fe_{II}$. a) switch off the NNN hoppings; b) switch on the NNN hoppings $t_3(Fe_{II}$- $Fe_{II})=0.4$, $t_4(Fe_{II}$ - $Si)=0.7$ (blue dashed lines show the parameters $t_1$ and $t_2$ from Table 1); c) switch on only the $Fe_{II}$ - $Fe_{II}$ hopping $t_3$

### 4. The spin-crossover.

Here we consider a possibility for the crossover from high-spin to low-spin states under hydrostatic pressure in $Fe_3Si$ [37,38]. $DO_3$ structure of $Fe_3Si$ has two types $Fe$-positions. The $Fe_I$ position is similar to the one in the ferromagnetic BCC-$Fe$; another iron $Fe_{II}$, has different from BCC-$Fe$ environment. Based on the $Fe_{II}$ pDOS shape the authors of [37] expected that such structure favors metamagnetic-like behavior under compression. The *ab initio* calculations of the total magnetic moment $\mu_{tot} = \mu_{FeI} + 2\mu_{FeII}$ of the cell of $Fe_3Si$ under pressure predicted existence of high- to low-spin crossover at the pressure $P$~50$kB$ [38] and $P$~150-200$kB$ [37]. The calculations of the magnetic moments of two inequivalent $Fe$ atoms has shown that it is the decrease of $\mu_{FeII}$ under pressure is responsible for the spin crossover; the $\mu_{FeI}$ almost does not dependent on the pressure.

The origin of the moments $\mu_{FeI}$, $\mu_{FeII}$ dependence on pressure $P$ can be understood within our model. Assuming that the hopping parameters depend on distance $\Delta R$ between the ions exponentially, $t = t_0 exp(\gamma \Delta R)$, where $t_0=t(P=0)$ and $\Delta R=R(P=0)-R(P)$. The equilibrium lattice parameters were determined from the *ab initio* GGA calculations via minimization of the enthalpy. Then, using the values of the hopping parameters at $P=0$ and at $P=250kB$, we obtained the following values for the parameters $\gamma$: $\gamma_1 = 0.68; \gamma_2 = 0.87; \gamma_3 = 1.31; \gamma_4 = 0.31$ for the parameters $t_1(P)$, $t_2(P)$, $t_3(P)$, $t_4(P)$ correspondingly.

[1]The calculation within our model with these parameters shows that such behavior of magnetic moments can be explained by their different dependence in the hopping matrix elements (see the upper panel of Fig.4) on pressure.

The pressure dependences of $\mu_{FeI}(P)$ in *ab initio* and model calculations are identical to each other, $\mu_{Fe_I}(P)$ weakly decreases with pressure (see the lower panel of Fig.4). The $Fe_{II}$ moment behavior in model differs from one in *ab initio* calculations in the region of high pressure $P > 300kB$: $\mu_{Fe_{II}}^{model}(P)$ decreased noticeably faster than $\mu_{Fe_{II}}^{GGA}(P)$, achieving the values $\mu_{Fe_{II}}^{model}(P = 500kB) = 0.27\mu_B$ whereas $\mu_{Fe_{II}}^{GGA}(P = 500kB)$ decreased only till 0.76$\mu_B$. Moreover, $\mu_{Fe_{II}}^{model}(P)$ experiences two jumps, at $P \approx 150k$ and $P \approx 400kB$. The decrease of the magnetic moment at $P \approx 150kB$ is in agreement with [37], however, they did not consider the region of higher pressures. We are not aware of the experiments, which confirm the existence of the high-spin to low-spin crossover in $Fe_3Si$, however, it was mentioned in the work [37] that the change of the absolute reflectivity near 100$kB$ was observed (reference [32] in [37]) in the reflectivity measurements on $Fe_3Si$ and further till the pressure up to 300$kB$ no changes were found. According to our model calculations one can expect the changes in optical properties in the interval of pressure 100 - 200$kB$, while the next peculiarities in moment behavior are expected at $P>400kB$.

---

[1] Notice, that such substantial changes of hopping parameters suggest that the often used first-order correction $\Delta t \sim \kappa \Delta P$ would be insufficient.

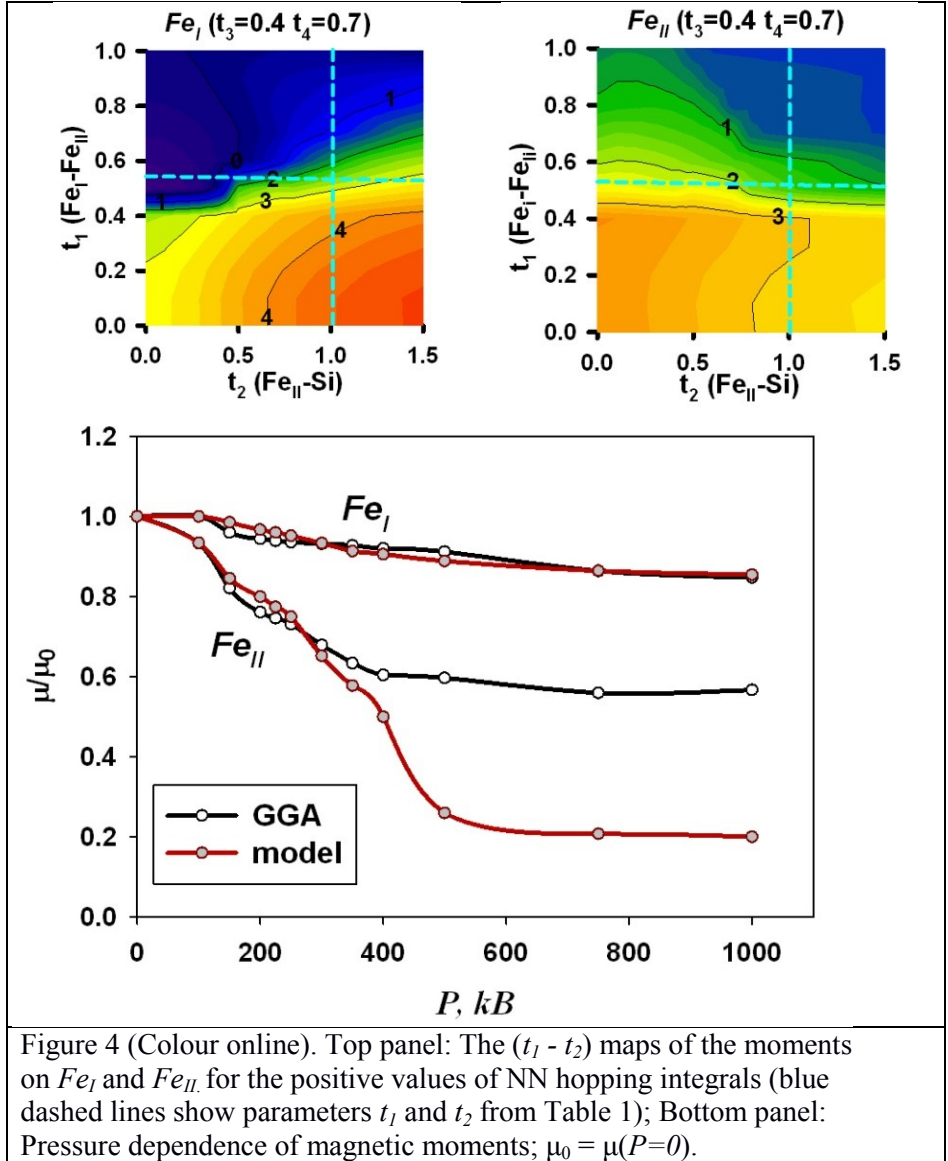

Figure 4 (Colour online). Top panel: The ($t_1$ - $t_2$) maps of the moments on $Fe_I$ and $Fe_{II}$ for the positive values of NN hopping integrals (blue dashed lines show parameters $t_1$ and $t_2$ from Table 1); Bottom panel: Pressure dependence of magnetic moments; $\mu_0 = \mu(P=0)$.

## 5. The alloys $Fe_xSi_{1-x}$

Let us now consider the dependence of magnetic moments in alloys $Fe_xSi_{1-x}$ on the concentration of non-magnetic atoms. The authors of [29] ascribed the changes of *Fe*-moment magnitude rather with an increase of number of NN non-magnetic atoms than with their concentration change. As well-known, in the $DO_3$ structure the non-magnetic atoms prefer to occupy the positions in the $Fe_I$ sublattice and form a partially ordered *CsCl*-like structure (B2). An increase of the *Si* concentration till 50%, i.e., when all $Fe_I$ atoms are replaced by *Si* ones (see Fig.5), the magnetic moment of the elementary cell vanishes. The partially ordered structure of the B2 type are observed in most of epitaxial *Fe - Si* films [17,39]. In amorphous films situation is different, there a chemically disordered random solid solution is formed: the *Si* atoms can occupy any of iron positions, $Fe_I$ or $Fe_{II}$. The electron and magnetic properties of the epitaxial and amorphous films are also essentially differ from each other [39,40]. Particularly, the magnetic moment per iron atom in amorphous films is higher, than in epitaxial [39,40]. For example, the experiments and *ab initio* calculations [39] show that in amorphous films the magnetic moment per *Fe* atom $\mu_{Fe} \approx 0.65\mu_B$ at 50% concentration of *Si*, contrary to the situation in the alloys with B2 structure. Thus, the possibility of occupation of the $Fe_{II}$ sites by *Si* atoms is one of main structural features which differ disordered random solid solution from B2-like structures and is directly related to the mechanisms of magnetic moment formation. As was described above, for the $Fe_3Si$ the critical role in moment formation on Fe atoms is played by the presence of Fe atoms in NNN sphere. Motivated by this fact we have considered hypothetical structures (see Fig.5b) where the additional *Si* atoms are placed into $Fe_{II}$ sublattice. In spite of the fact that both structures shown at Fig.5 contain the same concentration of *Si* atoms ($x$=50%), the moments on *Fe* atoms in these structures are very different. This difference is caused by the different local NN and NNN surrounding of Fe atoms. In first case, Fig.5a, at 50% *Si* concentration NN sphere of *Fe* contains only *Si* atoms, while the NNN one contains only iron. In the other structure, Fig.5b, the *Fe* atoms are absent in the NNN sphere of *Fe*. The dependencies of *Fe* moments on the hopping constants in these two structures are shown in the last column of Fig.5. The values of hoppings in *Fe₃Si*, $t_2$=1.0, $t_3$=0.4, are on the border between FM and PM states in the alloy with B2 structure. Notice that such a sharp border is characteristic for the moment maps of the *Fe* atoms which are surrounded by other *Fe* atoms (see Fig.5a and Fig.3c). A slight increase of the hopping $t_3$ stabilizes the PM state. Namely this mechanism works in B2 structure, where the equilibrium lattice parameter $a$(B2)=5.52Å < $a$(*Fe₃Si*). It is important that the PM state arises at NNN $t_3$≠0 only. The model with NN hoppings only, even if all NN to *Fe* atoms are *Si* atoms, does not have the solutions with zero moments on *Fe*. This statement contradicts to the conclusions of earlier (much less detailed) models of local environment, where the decrease of the moment on *Fe* atoms was ascribed to the increase of number of *Si* in NN sphere.

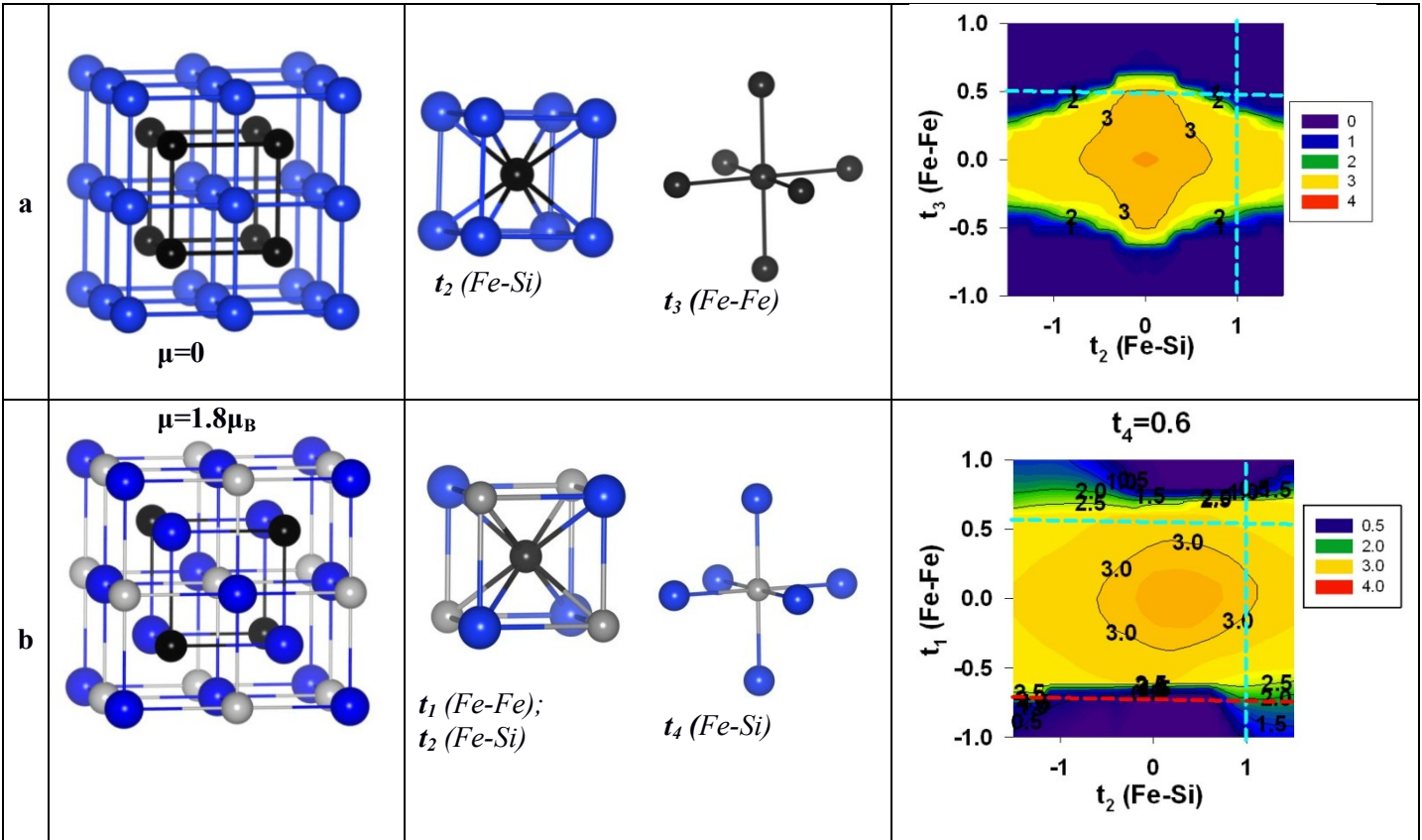

Figure 5 (Colour online). *Fe$_x$Si$_{1-x}$* alloys ($x$=50%): a) *CsCl*-like structure (B2), the types of NN and NNN hoppings and *t*-maps; b) hypothetical structure with *Si* atoms on the *Fe$_{II}$* sites (see text), the types of NN and NNN hoppings and *t*-maps. Blue balls are *Si* atoms, grey and black balls are *Fe* atoms. Blue dashed lines show parameters $t_1$=0.55 and t2=1.0, red dashed lines show parameter $t_1$=-0.75

The weak dependence of the *Fe* magnetic moment on the hopping $t_2$ between NN *Fe* and *Si* atoms is seen in all other cases: all the $t_1$ - $t_2$-maps for *Fe* moments, calculated within this model, are elongated along the axis $t_2$. The absence of *Fe* atoms in NNN sphere (Fig.5b) leads to two effects: i) the equilibrium lattice parameter increases till $a$=5.71Å and ii) the magnetic moment $\mu_{Fe} \approx 3\mu_B$ at *Fe* atoms arises at $t_1$=0.55 and $t_2$=1.0 (Fig.5b, blue lines), the values corresponding to *Fe₃Si*. The existence of NN *Si* - *Si* pairs distinguishes this structure from the DO₃ and B2 types of structure. The solution, corresponding to the *ab initio* calculations with $\mu_{Fe} \approx 1.8\mu_B$ arises in the model only at different signs of the NN hopping parameters, $t_1$ = –0.75 and $t_2$ = –0.9 (Fig.5b, red line).

The strong dependence of the magnetic moment on *Fe* atoms on the composition of NNN shell in the BCC type of alloys, possibly, is caused by particular spatial positions: NN atoms in BCC-like structures are arranged along the (111) direction, whereas NNN atoms are along (100). The strong σ-bond is formed from the *Fe* $e_g$-orbitals along (100) that delocalizes $e_g$ -electrons along this bond and, correspondingly, leads to a decrease of the magnetic moment.

## 6. Discussion and conclusions

The interpretation of the experiments [9-13] requires understanding of the role played by the local environment in the moment formation in iron silicides. This raises a question about the choice of a suitable theoretical tool for this analysis. The *ab initio* density-functional based methods allow for detailed description of the properties of compounds but it is difficult to extract from these calculations the contributions from the local environment. The many-body-theory based models allow to analyze the role played by different interactions in formation of the matter properties, as well as reveal the features, general for different classes of solids. However, these two approaches use different language. The methods, combining these two approaches, like LDA+U, LDA+DMFT, etc., contain the poorly controlled double-counting of intra-atomic interactions. The GW method requires so much of computer resources that the problems of solids with many atoms in elementary cell become inaccessible. It is possible to translate the results obtained within DFT approach to the language of many-body theory with the help of mapping the GGA-to-DFT results to a model. In our case the choice of the model was dictated by the facts that i) the delocalized *d*-electrons are responsible for a magnetism in the materials of interest, and ii) the intratomic Coulomb interactions are the largest matrix elements for them. This lead us to the multiorbital model with intraatomic interactions between d-electrons (similar to Kanamori model [36]), and *d-d*-intersite exchange interaction between delocalized *d*-electrons. The way how we did the mapping, as far as we aware, have not been used before. The leading idea is as follows. Since the *ab initio* methods treat the strongest part of the Coulomb interactions correctly via finding the best *self-consistent* charge density, minimizing the total energy, we choose the parameters of the model from the requirement that the model charge densities ( as well as density of electron states), obtained *self-consistently*, have to be as close as possible to the GGA ones. The model is solved within the Hartree-Fock approximation (HFA). The band structure arises due to hopping parameters, which connect nearest neighbors (NN) and next NN (NNN) sites. The HFA calculations show that the formation of moments on iron-atoms' is very sensitive to the values of namely NNN hopping parameters. This statement was demonstrated by comparison of the maps of moments' dependence on hopping parameters with and without taking into account NNN hopping parameters. This conclusion is especially interesting since most of models do not take the NNN hoppings into account.

The characteristic feature of these maps is presence of the regions with very fast change of magnetic moments as a function of hopping between two types of iron ($Fe_I$ - $Fe_{II}$); the NNN hopping ($Fe_{II}$ - $Fe_{II}$) makes the ferromagnetic (FM) region narrower and at large enough value causes the transition into ferrimagnetic state. Notice that the Stoner's like criteria for FM instability, which is natural for the models with delocalized electrons, is not sufficient here. Indeed, speculating from the strong-coupling-side perturbation theory (SCPT) the Hubbard-model-like effective antiferromagnetic interaction $-t^2/U$ forms the ferrimagnetic state (FiM; the state with oppositely directed but modulo different moments on inequivalent *Fe* atoms). So, the stabilizing FM state direct *Fe-Fe* exchange interaction should be strong enough to overcome the AFM-like contributions. In order to make mapping we had to use the weak-coupling theory (WCPT), however, the WCPT HFA captures the effects of SCPT the better the stronger is an orbital polarization. So, as seen from solutions, WCPT HFA is capable to take into account AFM interactions.

It may seem that the presented analysis is of theoretical interest only. However, the hopping parameters are the most sensitive parameters to different type of pressure. This statement follows from the fitting of hopping parameters of the model within the same scheme to the results of *ab initio* calculation of the enthalpy for $Fe_3Si$ at different pressures. The latter can be made negative, either chemically or by depositing the films on the substrate with larger than the iron silicide lattice parameter; or positive, by applying the hydrostatic pressure or depositing the films on the substrate with smaller lattice parameter.

The effect of the high-spin to low-spin crossover is predicted by both GGA and model calculations. Namely, the moment of $Fe_{II}$ atom sharply decreases whereas the moment of $Fe_I$ remains almost intact. The model calculation predicts also second crossover at higher pressure. It worth noting that here the spin-crossover arises not due to the standard mechanism of competition between the crystal-field splitting and *d-d* Hund exchange interaction, but is caused mainly by the delocalization of *d*-electrons.

At last, we have considered the mechanisms of moment formation in solid solutions $Fe_xSi_{1-x}$. Theoretically such alloys usually are studied within the coherent-potential approximation (CPA). By calculation of several hypothetical structures we explicitly show that *at the same concentration different magnetic structures arise* due to different NN and NNN environments for *Fe* atoms, the statement which is beyond reach of the CPA. Particularly, this finding allows also to explain the difference in the properties partially ordered alloys with B2 structures and the amorphous alloys.

Thus, we can conclude that, the decisive role in the formation of magnetic moments on *Fe* atoms is played by the effects of local environment in spite of the delocalized nature of d-electrons in the iron silicides. The contribution of NNN hoppings to it is far from to be negligible quite significant. The existence of the region with sharp transition from ferro- to paramagnetic state as well as the predicted spin crossovers strongly improve the perspectives of the practical applications of iron silicide films and, hopefully, will stimulate technologists to find a way to make the films near the instability line with desirable characteristics.

**Acknowledgement**

This work was supported by the Russian Foundation for Basic Research, projects № 14-02-00186, №16-42-240471, and by the Grants of the President of the Russian Federation for Support of Leading Scientific Schools № NSH-924.2014.2, № NSH-7559.2016.2. The authors would like to thank AS Shinkorenko for the technical support.

**References**
[1] Y. Maeda, T. Jonishi, K. Narumi, Y. Ando, K. Ueda, M. Kumano, T. Sadoh, M. Miyao, *Appl. Phys. Lett.* **91**, 171910 (2007)
[2] J. Kudrnovsky, N.E. Christensen, O.K. Andersen, *Phys. Rev. B* **43**, 5924 (1991)
[3] A. Bansil, S. Kaprzyk, P.E. Mijnarends, J. Tobola, *Phys. Rev. B* **60** 13396 (1999)
[4] K. Hamaya, K. Ueda, Y. Kishi, Y. Ando, T. Sadoh, M. Miyao, *Appl. Phys. Lett.* **93** 132117 (2008)
[5] K. Hamaya, T. Murakami, S. Yamada, K. Mibu, M. Miyao, *Phys. Rev. B* **83** 144411 (2011)
[6] O. Kubaschewski, *Iron-Binary Phase Diagrams* (New York: Springer-Verlag, 1982)
[7] G. Bertotti, A.R. Ferchmin, E. Fiorillo, K. Fukamichi, K. Kobe, and S. Roth, *Magnetic allows for technical applications. Soft Magnetic Allows. Invar and Elinvar Allows* (Landolt Bornstein, New Series vol.III/19il), (Berlin: Springer) pp 33-142 (1994)
[8] V.A. Niculescu, T.J. Burch, J.I. Budnick, *JMMM*, **39**, 223-267 (1983)
[9] S.Yoon and J.G.Booth, *Phys.Rev.Lett.* **48A**, 381, (1974)
[10] A. Paoletti and L. Passari, *Nuovo Cimento* **32**, 25 (1964)
[11] A.K. Arzhnikov, L. V. Dobysheva, E.P. Yelsukov, G. N. Konygin, E. V. Voronina, *Phys. Rev. B* **65**, 024419 (2002)
[12] T.J.Burch, T.Litrenta, J.I.Budnick *Phys.Rev.Lett.* **33**, 421 (1974)
[13] M.B.Stearns *Phys.Rev.B* **4** 4069 (1971)
[14] D. Berling, G. Gewinner, M.C. Hanf, K. Hricovini, S. Hong, et. al *JMMM*. **191**, 331 (1999)
[15] A.Ionescu, C.A. Vaz,T.Trypiniotis, C.M.Curtler, H.Garsia-Miquel,J.A.C.Bland *Phys.Rev.B* **71**, 094401 (2005)
[16] M. Walterfang, W. Keune, K. Trounov, R. Peters, U. Ruecker, K. Westerholt, *Phys.Rev. B* **73**, 214423 (2006)
[17] A. X. Gray, J. Karel, J. Min´ar, C. Bordel, et. al. *Phys. Rev. B* **83**, 195112 (2011)
[18] Yakovlev I.A., Varnakov S.N., Belyaev B.A., et.al, *JETP Letters. 2014. Vol. 99, Is. 9. P. 527-530 (2014)*.
[19] Lyaschenko S.A., Popov Z.I., Varnakov S.N. et.al, *JETP, Vol. 120, No. 5,pp.886-893.(2015)*
[20] Guixin Cao, D.J.Singh, X.-G.Zhang et.al. Phys.Rev.Lett. **114**, 147202 (2015)
[21] W.A.Hines, A.H.Menotti, J.I.Budnick,T.J.Burch,T.Litrenta,V.Niculescu, K.Ray *Phys.Rev. B* 13,4060 (1976)
[22] E P Elsukov, G N Konygin, V A Barinov: and E V Voronina *J. Phys.: Condens. Matter* **4,** 7597 (1992)
[23] J.Kudrnovsky, N.E.Christensen, O.K.Andersen *Phys.Rev. B* **43**,5924, 1990
[24] T. Khmelevska, S.Khmelevskyi, A. V. Ruban and P. Mohn *J. Phys.: Condens. Matter* **18,** 6677 (2006)
[25] N.I.Kulikov, D.Fristol, J.Hugel, A.V.Postnikov *Phys.Rev. B* **66**,014206 (2002)
[26] A.Bansil, S.Kaprzyk,P.E.Mijnsrends, J.Tobola *Phys.Rev. B* **60**, 13396 (1999)
[27] A. Go, M. Pugaczowa-Michalska, and L. Dobrzyrnski *Eur. Phys. J. B* **59**, 1 (2007)
[28] E. Voronina et al. *Nuclear Instruments and Methods in Physics Research A* **575,** 189 (2007)
[29] A.K. Arzhnikov, L. V. Dobysheva *Itinerant Electron Magnetism: Fluctuation Effects* Volume 55 of the series NATO Science Series pp 375-389 (1998)
[30] E.J.Garba and R.L.Jacobs *J.Phys.F: Met.Phys.* **16,** 1485 (1986)
[31] J.C. Slater, G.F.Koster *Phys.Rev.* **94**,1498 (1954)
[32] G.Kresse and J. Furthmuller *Comput. Mat. Sci.* **6**, 15 (1996); G.Kresse and J. Furthmuller *Phys. Rev.B* **54** 11169 (1996)
[33] P.E.Blochl *Phys. Rev.B* **50**, 17953 (1994); G.Kresse and D.Joubert *Phys. Rev. B* **59**, 1758 (1999)
[34] J.P.Perdew, K.Burke and M.Ernzerhof *Phys.Rev.Lett.* **77**, 3865 (1996); J.P.Perdew, K.Burke and M.Ernzerhof *Phys.Rev.Lett*. **78**, 1396 (1997)
[35] H.J.Monkhorst and J.D.Pack *Phys.Rev.B* **13**,5188 (1976)
[36] J. Kanamori *Prog. Theor. Phys*. **30**, 275 (1963)

[37] N.E.Christensen, J.Kudrnovsky, C.O.Rodrigues *Int. J.Mater.Sci.Simu*. **1**, 1 (2007)

[38] Joo Yull Rhee, B.N.Harmon *Phys.Rev. B* **70**,094411 (2004)

[39] Karel, J. Juraszek, J. Minar, C. Bordel, K et. al. *Phys. Rev. B* **91**, 144402 (2015

[40] J Karel, Y N Zhang, C Bordel, K H Stone, et. al. *Mater. Res. Express* **1,** 026102 (2014)